\documentclass[twocolumn]{aastex631}

\begin{document}

\title{Investigating the Formation of Planets Interior to \emph{in situ} Hot Jupiters}

\author[0009-0009-0948-8819]{Devansh Mathur}
\affiliation{University of Wisconsin - Madison Department of Astronomy, 475 N. Charter St. Madison, WI 53706, USA}
\author[0000-0002-7733-4522]{Juliette Becker}
\affiliation{University of Wisconsin - Madison Department of Astronomy, 475 N. Charter St. Madison, WI 53706, USA}

\begin{abstract}
The population of hot Jupiters with adjacent planetary companions is small but growing, and inner companions appear to be a nearly ubiquitous outcome within this subset of the exoplanet census. While most hot Jupiters are believed to form via tidal migration, the presence of adjacent companions is not easily explained by this formation mechanism, requiring consideration of additional formation mechanisms such as disk migration and in situ formation. In this work, we explore the possibility of in situ formation for both hot Jupiters and their interior companions. Using numerical simulations performed with the N-body integrator {\tt\string REBOUND}, we investigate the growth of interior companions under various assumptions about disk conditions and hot Jupiter final orbital positions. Our results show that if a sufficiently high density of planetary embryos is transported to short orbital radii, it is feasible for both hot Jupiters and their interior companions to form in situ, providing a viable explanation for a subset of observed planetary architectures.
\end{abstract}

\keywords{Exoplanets (498) --- Planet Formation (1241)--- Planetary Migration (2206) --- Hot Jupiters (753) --- Planetesimals (1259)}

\section{Introduction} \label{sec:intro}
Planet formation is a foundational topic in astronomy, yet the rapidly growing diversity of observed exoplanetary systems continues to challenge prior theories developed from our Solar System. The discovery of hot Jupiters in 1995 \citep{Mayor1995} posed a significant challenge to classical models of planet formation, particularly regarding the presence of giant planets at small orbital separations \citep[see a review in][]{DawsonJohnson2018}. These findings prompted a reevaluation of planetary system architectures and the processes that could give rise to such configurations \citep{Lin1996, Rasio1996}, as Jupiter-mass planets were  thought to form in the outer parts of protoplanetary disks \citep{Pollack1996}.

\begin{figure}[ht!]
    \centering
    \includegraphics[width=1\linewidth]{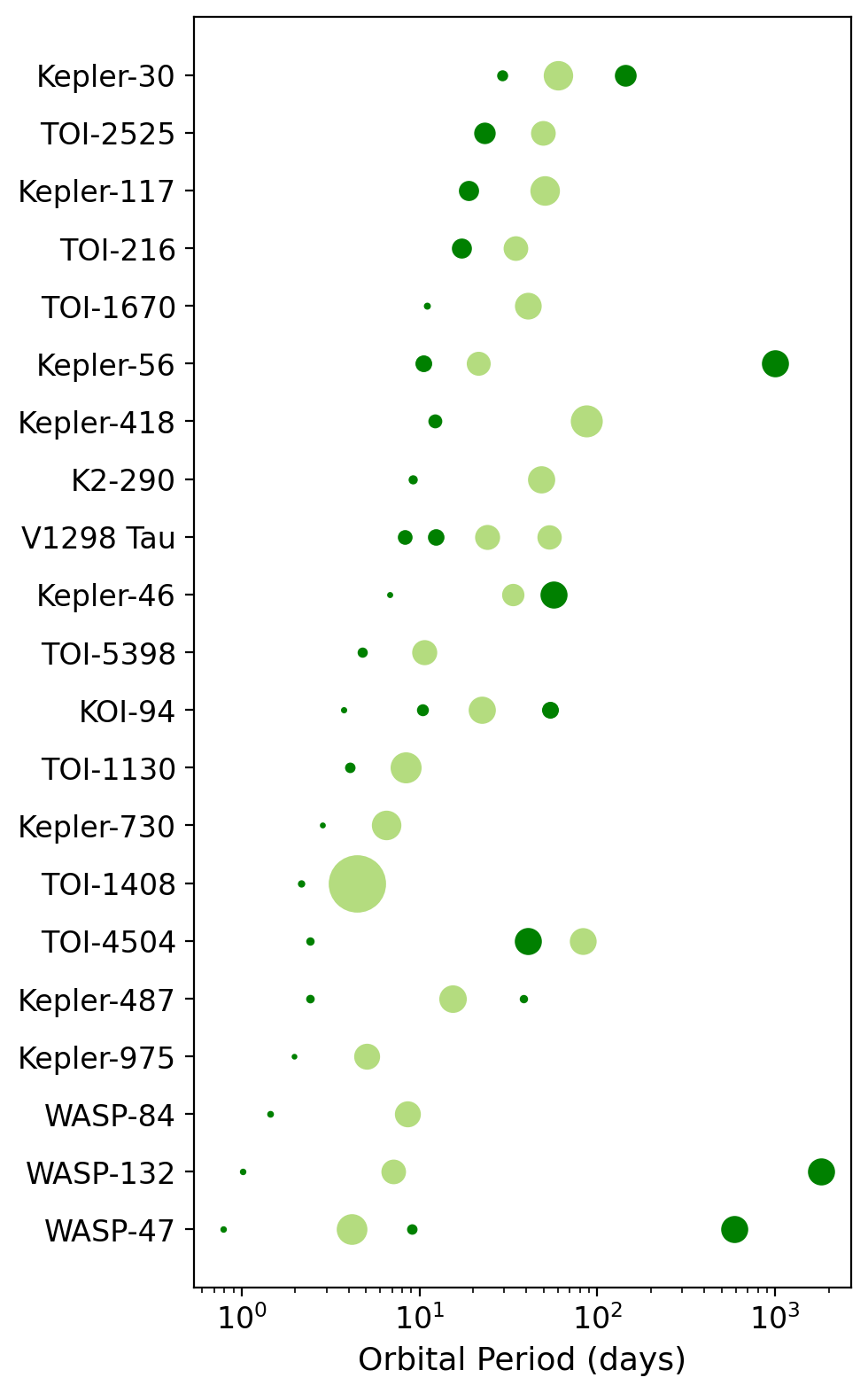}
    \caption{Systems selected from the IPAC Exoplanet Archive \citep{Christiansen2025} that contain a hot (orbital period $P\le10$ days) or warm Jupiter (orbital period $10<P<100$ days) and additional planets within $P<100$ days. Marker sizes are scaled by planet radius. Hot/warm Jupiters are denoted by light green circles (and identified as those meeting the above orbital period definitions and with radius $R_p> 0.8 R_J$), and smaller planets or cold Jupiters are denoted by dark green circles. }
    \label{fig:multiplanet_systems}
\end{figure}

The in situ model for hot Jupiter formation has long been considered as a potential explanation for the existence of gas giant planets at short orbital periods \citep{Bodenheimer2000}. 
In situ formation models refer to those where Jupiter-mass planets accreted their envelopes while residing on short-period orbits \citep{Boley2016, Batygin2016}. 
One of the primary challenges to this model is the limited gas budget available close to the star. However, previous studies have shown that if a super-Earth core can form early enough, runaway gas accretion may proceed efficiently even at short separations \citep{Lee2014}, making the in situ formation of hot Jupiters feasible.

The in situ model can also be invoked as a solution for another hot Jupiter puzzle: adjacent companions to hot and warm Jupiters. While evidence suggests that tidal migration is the dominant mechanism that produces hot Jupiters \citep{Petrovich2016, Zink2023}, tidal migration models \citep[e.g.,][]{Fabrycky2007} often destabilize nearby companions due to the eccentricity evolution of the proto-hot-Jupiter. In situ formation or disk-driven migration of hot Jupiters could allow for the survival of adjacent planets. This makes in situ formation an attractive origin theory for hot Jupiters that have nearby planetary companions, particularly for those with {both interior and exterior companions} \citep[e.g.,][]{Becker2015}.

However, an additional challenge emerges for the in situ scenario: all currently published hot Jupiters with adjacent planetary companions have companions interior to the hot Jupiter \citep[e.g.,][]{Canas2019,Huang2020, Hord2022, Sha2023, Maciejewski2023, Korth_2024}, a geometry which is also common in warm Jupiter systems.
The ubiquity of these inner companions {amongst the sample of hot Jupiters with any nearby companions (which as a population is a small subset of all hot Jupiters)}, as seen in Figure \ref{fig:multiplanet_systems}, is not surprising {for two reasons: 1) inner companions have a higher transit probability than outer companions{, and as a result inclination dispersion within a system is more likely to decrease the co-transit probability of an outer companion to a hot Jupiter than of an inner companion}; and 2) }if hot Jupiters formed in situ {(before the onset of stellar quadrupole-driven secular resonances, e.g., \citealt{Li2020, Brefka2021})} and had detectable inner companions, those companions would {be more} likely {to} remain in the same orbital plane as the hot Jupiter {than outer companions}, regardless of the host star’s obliquity or other dynamical influences
\citep{Spalding2014, MacLean2025}.

In this study, we investigate the extent to which the in situ case is feasible through a suite of N-body simulations. {We build upon the work of \citet{Poon2021}, who investigated this problem during the gas disk phase, by considering the accretion of planetary embryos in the gas-free disk.} In Section~\ref{sec:methods}, we describe the numerical setup, including initial conditions, collision handling, and the range of parameter space explored. Section~\ref{sec:results} presents the outcomes of our simulations, and interpretation of the implications of our results towards the feasibility of in situ formation as a pathway to assemble hot Jupiters with interior companions. In Section~\ref{sec:discussion}, we evaluate the feasibility of in situ formation based on our simulation results and discuss caveats and future directions for further work. We conclude in Section~\ref{sec:conclusion} with a summary of our main results.

\section{Methods} \label{sec:methods}

\subsection{Dynamical Motivation}

The surface density of solids in a protoplanetary disk can be written as \citep{Hayashi1981}:
\begin{equation}
    \Sigma_Z(r) = \Sigma_0 \left( \frac{r}{1\,\text{au}} \right)^{-p}
\end{equation}
where $\Sigma_Z(r)$ is the surface density of solids \citep[denoted by $Z$ following the notation of][]{Hansen2012} at orbital distance $r$, $\Sigma_0$ denotes the surface density at $r = 1$ au, and $p$ is the power-law index of the disk surface density profile. For the typical Minimum-Mass Solar Nebula {(MMSN)}, $\Sigma_0 = 7$ g/cm$^2$ \citep{Hayashi1981, Hansen2012}.


Historically, these surface density profiles were interpreted to imply that large planets cannot form at short orbital distances, simply because there is not enough solid material available (with less than 3 $M_{\oplus}$ being available within 0.5 au for even the more favorable model). However, observations from \textit{Kepler} and \textit{TESS} \citep{Borucki2010, Borucki2011, Howell2014, Guerrero2021} clearly demonstrate that the dominant outcome of planet formation is the emergence of multiple super-Earths within about 0.5 au \citep{Weiss2023, Howe2025}. \citet{Chiang2013} highlighted this discrepancy and proposed the `maximum-mass extrasolar nebula' (MMEN) — a modified surface density profile required to account for the observed planets forming at their current locations. {More recent work by \citet{Dai2020} evaluated the dependence of the MMEN on stellar parameters and found that the MMEN is roughly an order of magnitude more dense than the MMSN.} {This discrepancy has motivated the possibility that processes beyond in situ formation, such as migration, have shaped these systems.}

This mismatch has led to the widespread view that planetary migration likely plays a key role in shaping the architecture of inner planetary systems \citep{Hansen2012}.  {Following the arguments of \citet{Hansen2012}, our working assumption in this work, based on the results of \citet{Boley2016} and \citet{Batygin2016}, is that the super-Earth–sized core of the hot Jupiter forms at moderate distances in the disk (within 0.1–2 AU) and then migrates inward. Only after this inward migration {has halted (due to, for example, migration traps near the inner edge of the protoplanetary disk, the exact locations of which will depend on disk and planet properties \citealt{Zawadzki2022})} does the hot Jupiter undergo accretion of its gaseous envelope. Throughout the manuscript, when we use the term `in situ formation', we are referring specifically to this envelope accretion stage, which we assume occurs close to the final orbital position of the planet.} 

\citet{Boley2016} suggest{s} that hot Jupiters could form in situ if one of these {inwardly migrated} super-Earths crosses the critical mass threshold to initiate runaway gas accretion. This requires relatively rapid assembly, as gas disks dissipate quickly \citep[within 3--10 Myr][]{} in concert with declining gas-to-dust ratios \citep[e.g.,][]{Birnstiel2010, Zhang2025, Trapman2025}. If hot Jupiters form via this pathway, beginning as super-Earth cores that form farther out and migrate inward before accumulating gas, then the migration process likely transports solids inward as well. Specifically, any material originally located interior to the core's formation site would be swept inward alongside it, potentially contributing to the formation of other planets.

The inward funneling of solid mass would result in an increased surface density of solid material. 
We assume in this work that the material being pushed inward was likely already in the form of planetary embryos: bodies large enough to exert gravitational influence under the right conditions but perhaps too dynamically isolated pre-migration to grow larger in their original orbits.

Short-scale convergent migration could change that. As these embryos are forced inward together, they may begin to gravitationally interact in ways that were not previously possible. These interactions could destabilize some embryos, ejecting them from the system, or drive collisions and mergers, processes that could lead to the formation of larger planets.

In the next section, we explore this question using numerical simulations, aiming to answer the following question: if hot Jupiters form through an in situ pathway, where the super-Earth sized core of the planet may have experienced some short-range migration pathway, could this process also give rise to detectable interior planetary companions, as seen in many hot Jupiter systems (Figure \ref{fig:multiplanet_systems})?

{Recent work by \citet{Poon2021} has explored the in situ formation of hot Jupiters and nearby super-Earth companions using sophisticated simulations that included gas accretion, eccentricity and inclination damping, and an evolving protoplanetary disk and a solids density of $10^3$ g/cm${2}$ within 0.5 AU. \citet{Poon2021} distributed their mass in embryos of 0.5 $M_{\oplus}$. Their inclusion of a gas disk allowed them to assess the rate of hot Jupiter formation, and they found that while forming a hot Jupiter is rare when starting from only planetary embryos (occurring about 1\% of the time), if a hot Jupiter forms, it is very likely to have a companion (95\% of the time). Our work builds on these results by focusing specifically on the late-stage, gas-free evolution of planetary embryos located interior to a hot Jupiter. For computational efficiency, we start with the assumption that a hot Jupiter has formed, and vary both the solid surface density and the orbital distance of the giant planet to assess whether companions forming with various configurations would be observable or not. }

\subsection{N-body Simulation Setup: Effect of Surface Density on Planet Formation}
In this section, we use a suite of N-body simulations to evaluate how varying the surface density of solid material affects the maximum planetary core mass that can be formed in close-in, gas-free environments. All simulations are conducted using the {\tt\string REBOUND} package \citep{Rein2012, Rein2017} with the TRACE integrator \citep{Hernandez2023, Lu2024}, which dynamically adjusts the timestep during integration and is designed to accurately and efficiently resolve close encounters between bodies.

Each simulation is initialized with a Sun-like central star with a mass $M_*=1\:M_{\odot} $ and a Jupiter sized planet with $M_p = 1 \: M_J$ on an almost circular orbit with $a = 0.05$ au and $e = 0.1$, consistent with a possible outcome of the in situ formation pathway. We assume that the hot Jupiter has already reached its final orbital position at the time our simulations begin, such that the surrounding disk has lost its gas component, but the solid material in the form of planetary embryos still remains.

We ran a total of {159 simulations}, each initialized with 30 planetary embryos distributed randomly interior to the hot Jupiter, between 0.01 and 0.04 au and {integrated for 1 Myr}. {In the final 100kyr of the simulations, 82\% of embryos show a fractional semi-major axis variation below 1\%, indicating that a 1 Myr integration captures the system's long-term dynamical state.} {This initial condition does not consider whether disk migration would capture the embryos into resonant orbits, although previous work \citep{Mandell2007} shows that this may happen. }

{The initial spacing between embryos is best characterized in units of mutual Hill radii. We use $\Delta$ to denote the number of mutual Hill radii between an adjacent pair of embryos. The mutual Hill radius can be computed as \citep{Kokubo1998}
\begin{equation}
    r_{\text{H,mut}} = \left( \frac{m_1+m_2}{3M_{\odot}} \right)^{1/3}  \frac{a_1+a_2}{2}
\end{equation}
Across all simulations, adjacent embryo-embryo separations range from $0 \lesssim\Delta\lesssim 300$ across all simulations, though the upper limit of this range is skewed by sparsely spaced, low-mass pairs in some cases. The median separation is $\Delta\approx 7$, indicating that most embryos are initially closely packed. For embryo-giant planet separations, we find a range $3.3\leq\Delta\leq19.5$ with a median of $\Delta \approx10.3$, such that the closest separations are at approximately the stability threshold defined by \citet{Chambers1996}.} While the initial timestep is set to 5\% of the innermost planet's orbital period, the TRACE integrator dynamically adjusts the timestep throughout the simulation, decreasing it during close encounters.
For each simulation, we varied the total solid mass within a range consistent with a large range of plausible surface densities, sampled between roughly $5\times10^{1}$ and $8\times10^{5}$ g/cm$^{2}$. 
To generate the planetary embryo masses for each simulation, we set bounds for the total amount of mass in the {embryo disk}, with the lower limit at $1 \: M_{\oplus}$ and the upper limit at $10 \: M_{\oplus}$. Between these endpoints, we generate 10 different values of total planetary embryo disk mass, spaced logarithmically to sample across several orders of magnitude. For each of these 10 total masses, we define a disk mass range by taking $\pm 50\%$ of the central value, resulting in a unique lower and upper bound for the total disk mass. 
To assign individual embryo masses, we draw 30 values from a uniform distribution between these bounds, then divide the drawn values by 30 so that the total mass of all 30 embryos sums to reside in the chosen disk mass range. 
Each embryo begins with with zero inclination ($i = 0^\circ$) and eccentricities randomly sampled from a uniform distribution in the range $0 \le e \le 0.01$. This coplanar setup reflects the expectation that prior gas damping during migration would suppress mutual inclinations among bodies and establish a dynamically cold configuration \citep{Tanaka2004}.

When planetary embryos are ejected from the system or collide with the central body, they are removed from the simulation. Collisions between the planetary embryos or between a planetary embryo and the hot Jupiter are treated as perfect mergers, conserving mass and linear momentum. 
Our chosen collision routine omits fragmentation. While fragmentation effects may change the timescales over which growth occurs \citep{Clement2019}, we expect that the qualitative results will not be affected \citep{Kokubo2002,Leinharadt2005} and significant fragmentation will not occur in our modeled gas-free environment \citep{Xie2008}. 
We check for collisions using the scaled radii of planetary embryos ($R_s$) using the mass-radius relationship of \citet{Seager2007}:
\begin{equation} \label{eq:radius_calc}
\log_{10} R_s = k_1 + \frac{1}{3} \log_{10} (M_s) - k_2 M_s^{k_3}.
\end{equation}
At extremely short orbital radii (such as the 0.01 - 0.04 au considered in this analysis), high surface temperatures can lead to the evaporation of silicates leading to a bulk composition of Fe($\alpha$) \citep{Cameron1985, Johansen2022}, so we choose $k_n$ values corresponding to Fe($\alpha$):  $k_1 = -0.20945\: , \: k_2 = 0.0804\: \:\text{and }\: k_3 = 0.394$ \citep{Seager2007}. 

\begin{figure}[htbp!]
    \centering
    \includegraphics[width=1\linewidth]{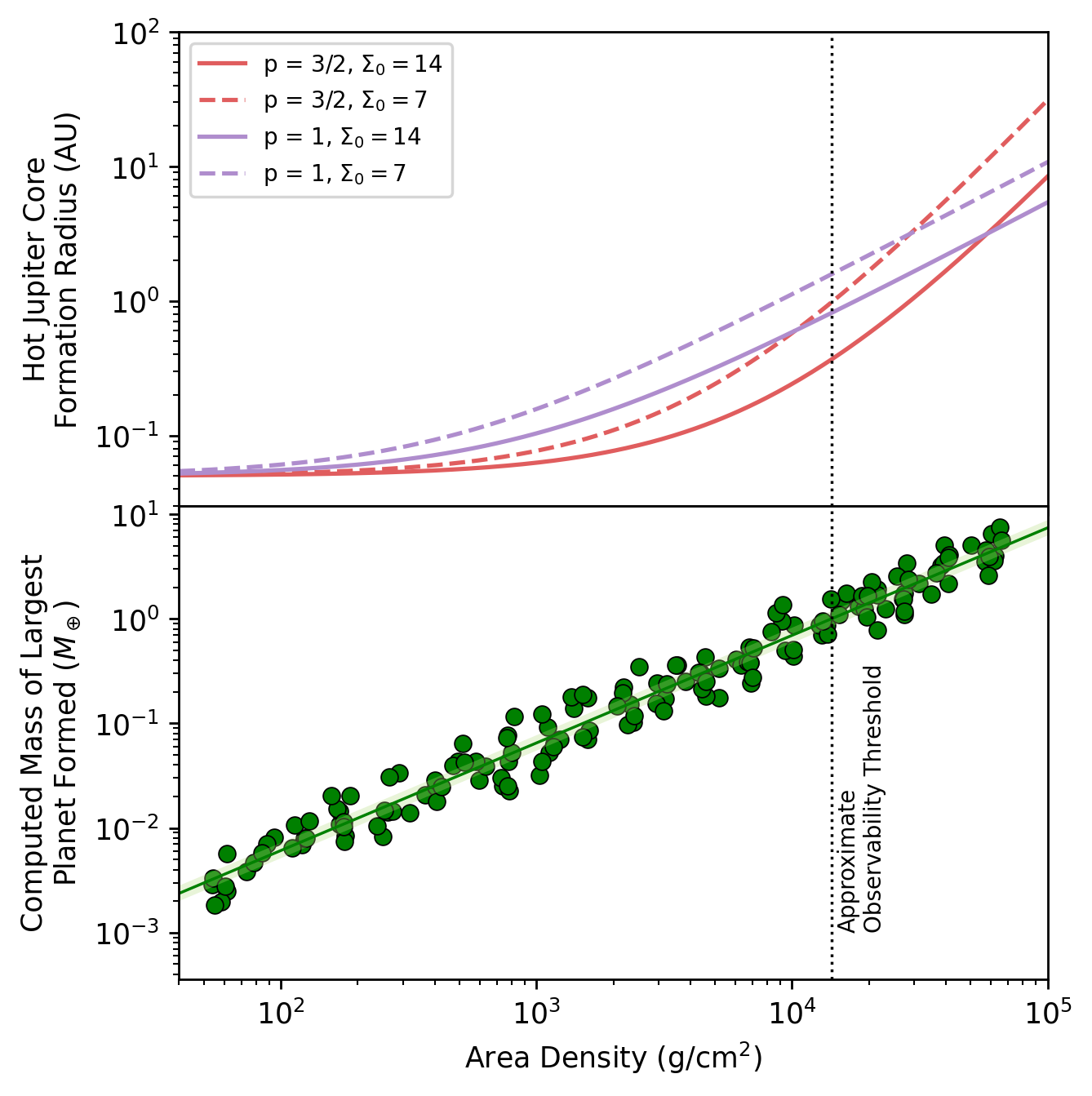}
    \caption{A comparison between the planet size expected from a given initial surface density of planetesimals and the orbital location at which a hot Jupiter's core would need to have formed in order to deliver that surface density interior to its final orbit after migration. \emph{Bottom panel:} The results of our numerical simulations showing the largest planet formed for a population of planetary embryos interior to the hot Jupiter with varying surface densities of solids. Each green dot represents the result from a unique simulation, and we overlay a best-fit model to the results to guide the eye. \emph{Top panel:} For the range of surface density of solids considered in our simulations, we show the inferred formation radius of the core of the hot Jupiter for four different disk models.}
    \label{fig:area_density_plot}
\end{figure}

The results of our initial suite of simulations are presented in Figure \ref{fig:area_density_plot}. In the bottom panel, each point corresponds to a simulation with a different total solid surface density but otherwise identical set-ups. The effective surface area density of planetary embryos, $\Sigma_p$, is computed as in \citet{Becker2017}:
\begin{equation} \label{eq:density_calc} \Sigma_p = \frac{1}{\pi (a_{out}^2 - a_{in}^2)} \sum_{i=1}^n m_i \end{equation}
where $m_i$ denotes the mass of each planetary embryo, the innermost planetary embryo has semi-major axis $a_{in}$, and the outermost planetary embryo has semi-major axis $a_{out}$.
As expected, we find a positive correlation: higher surface densities result in larger objects formed via the collisional accretion of planetary embryos.

In the top panel of Figure \ref{fig:area_density_plot}, we show the inferred proto-hot-Jupiter core formation radius for various disk parameters. 
{In interpreting the top panel of Figure \ref{fig:area_density_plot}, we adopt two simplifying assumptions. First, we assume that the assembly of planetary embryos is perfectly efficient, such that all available solids are incorporated into planets without any losses \citep[which is likely an overestimate;][]{Mandell2007}. Second, we assume that, by the time migration begins, all solid material interior to the hot Jupiter’s core formation radius is already contained in the form of planetary embryos. Implicit in this framework is the further assumption that the hot Jupiter core formation radius defines the boundary of the region from which interior material can be assembled into planets \citep[i.e., no disk material can cross the orbit of the migrating proto-hot Jupiter, which may not be true for all geometries;][]{VanClepper2025}.
Under these idealized conditions, the top panel of Figure \ref{fig:area_density_plot} highlights how variations in $\Sigma_0$ and $p$ determine the location of core assembly, and therefore which solids are available for building additional planets.}

For comparison, we also include in Figure \ref{fig:area_density_plot} an approximate detectability threshold of $1 \ \text{M}_\oplus$, which marks the observational limit below which planets are unlikely to be detected in current transit surveys. This value is just a threshold to aid in the interpretation of our results, as the actual detectability depends on stellar brightness, photometric variability, and observational baseline \citep{Gaudi2005, Gaudi2005size}, all of which may vary widely between targets being observed.


\subsection{Additional N-body Simulations: varying the formation location of the hot Jupiter}\label{spacing_sims}
In the previous section, we varied the surface density of planetary embryos while keeping the orbital distance of the hot Jupiter fixed. However, the orbital radius of the hot Jupiter will drive the scale of evolution for nearby planetary embryos. Gravitational interactions between the hot Jupiter and the inner embryo disk can excite eccentricities and relative velocities in the disk, potentially leading to mergers, ejections, or destabilization of otherwise stable orbits.
If the hot Jupiter resides at a large orbital distance, its influence on the inner embryo disk may be negligible, allowing embryos to evolve largely unaffected. Conversely, if the hot Jupiter is extremely close to the star, its gravitational perturbations on nearby embryos can be more severe.

To examine this effect, we performed an additional suite of simulations in which we fix the {surface density of the embryo disk} and instead vary the orbital radius of the hot Jupiter. 
In this second suite of simulation, we vary the hot Jupiter’s semi-major axis between $0.03 \text{ au}\le a\le 0.10$ au \citep[with these limits inspired by][]{Beaug2012,Udry2007}.
{We ran {164} total simulations for 1 Myr each}. In each simulation, the semi-major axis of the hot Jupiter was randomly sampled from this range. {To isolate the effect of the orbital radius of the hot Jupiter, we fixed the surface density across all runs at $2\times10^4\ \text{g/cm}^2$, which is approximately our observability threshold as computed in the previous section (vertical line in Figure \ref{fig:area_density_plot}). This surface density was then used to assign masses to the embryos. At the lower end of the tested range for the hot Jupiter's semi-major axis (0.03 au), embryos have masses of $2.4 \times 10^{-2}\ \text{M}_{\oplus}$, while at 0.1 au they are initialized at $0.63\text{M}_{\oplus}$. While the masses at the upper end of the range are physically unrealistic for objects we would consider embryos, using a fixed surface density controls against the variations in collision history caused by variations in surface density (Figure \ref{fig:area_density_plot}).} 

\begin{figure}[htbp!]
    \centering
    \includegraphics[width=1\linewidth]{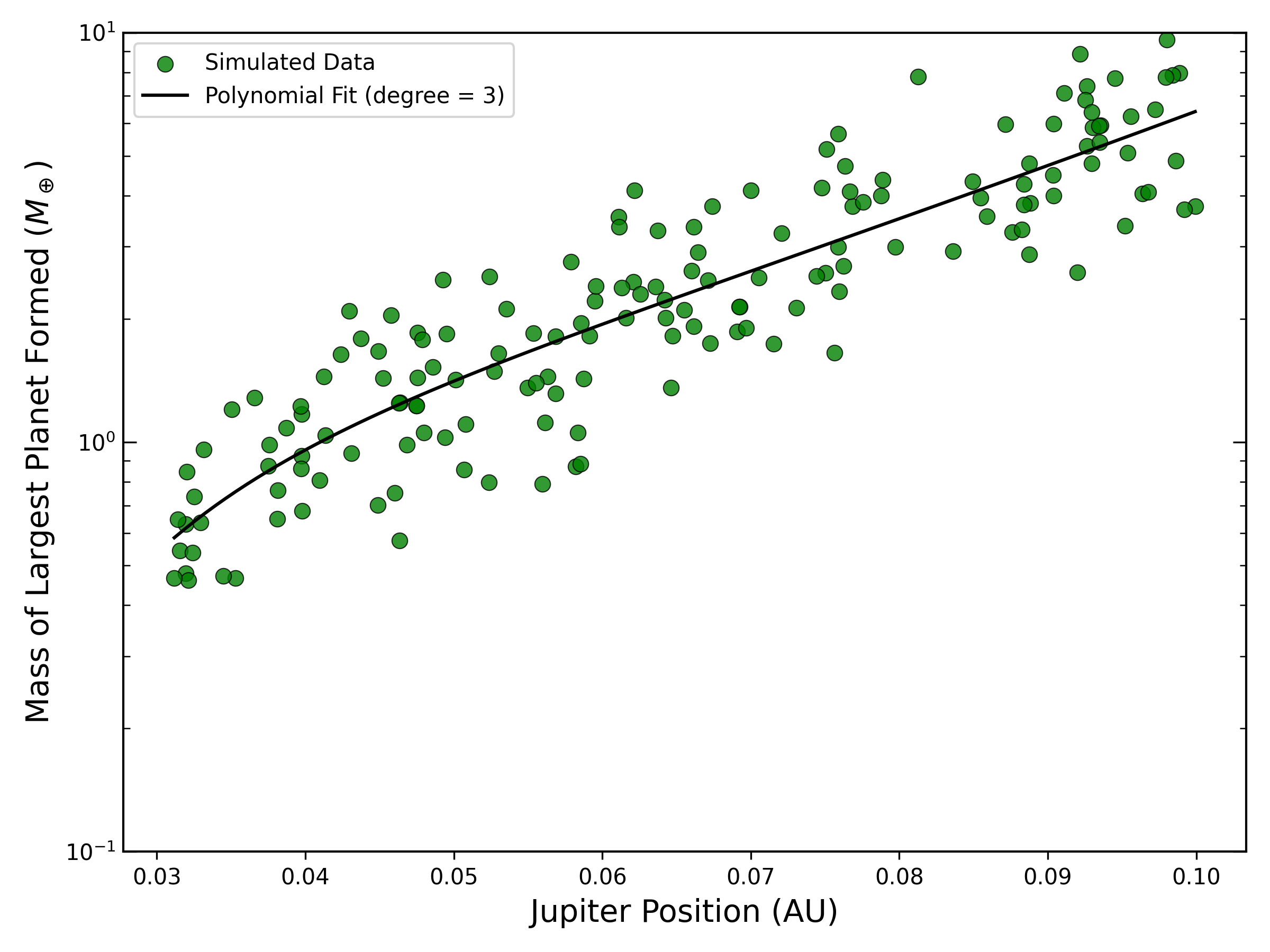}
    \caption{{Mass of the largest planet formed as a function of the hot Jupiter's orbital position. Green points indicate simulation outcomes, and the black curve shows a polynomial fit to guide the eye.}}
    \label{fig:spacing_plot}
\end{figure}

This second simulation suite allows us to isolate how proximity to the giant planet affects planet growth and stability in an otherwise identical simulation setup to that used in the previous section. 
The results of this second suite of simulation is shown in Figure \ref{fig:spacing_plot}, where we show the largest planet formed via accretion for each tested value of hot Jupiter orbital radius.

\section{Results} \label{sec:results}
\subsection{Effect of Planetary Embryo Area Density}
The results from our first suite of simulation are shown in the bottom panel of Figure~\ref{fig:area_density_plot}. These exhibit a strong positive correlation across several orders of magnitude in surface area density, with the mass of the most massive planet formed ranging between around 0.01 $M_\oplus$ at the lowest disk surface densities to nearly 10 $M_\oplus$ at the highest. {We also find that on average $\sim75-80\%$ of the initial solid mass remains in the planets formed by the end of the integration. The remaining $\sim20-25\%$ is lost from the system primarily through accretion onto the central star.}
This is fully consistent with theoretical expectations that higher local mass densities increase the rate and efficiency of collisional growth \citep{Kokubo1998}. The top panel of Figure \ref{fig:area_density_plot} shows the orbital radius at which the core of the hot Jupiter would have needed to form to create each surface density, under the assumption that all solid material interior to the core's formation radius migrated inwards and formed a disk of planetary embryos between 0.01 and 0.04 au. These values can be considered lower limits for the core formation radius, as any loss of disk material (via ejections or other mechanism) would act to decease the final surface density. 

Based on the surface density of planetary embryos required to form a planet of at least one Earth mass, we infer that the core of the hot Jupiter must have originated at minimum orbital distances of at least 0.4 - 1.6 au, depending on the disk models. For the four disk models under consideration, we find that if all solid material interior to 0.4, 0.8, 1.0, and 1.6 au had formed the disk of planetary embryos, an observable interior companion could have been accreted based on the results of our simulations. 
For the disk models considered in this work, this implies that, in order for in situ formation to be viable, the super-Earth core that eventually accreted gas to become a hot Jupiter must have formed exterior to those radii.

Consequently, for systems where hot Jupiters are observed to have interior companions, this suggests that the hot Jupiter likely underwent some degree of short-scale disk migration. However, in the case of heavier disks with higher surface density normalizations or steeper profiles, the required migration distance may be relatively modest, on the order of as little as 0.4 au rather than 1-10 au. This supports the plausibility of in situ hot Jupiter formation via short-range migration combined with core accretion under favorable disk conditions for hot Jupiters with interior planetary companions.

\begin{figure}[htbp!]
    \centering
    \includegraphics[width=1\linewidth]{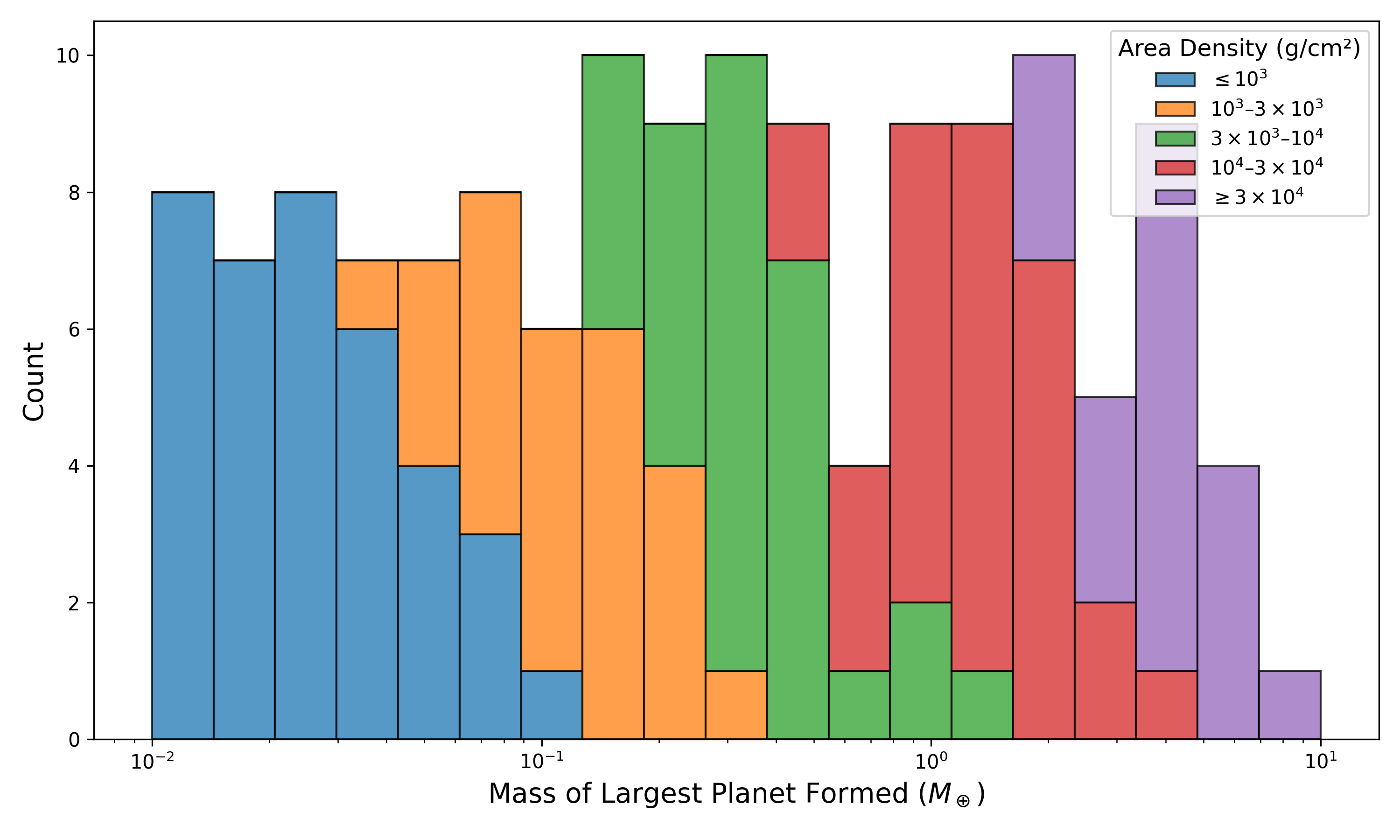}
    \caption{Frequency distribution of the most massive planet formed in each simulation from our first suite of simulations. Each color represents a bin of disk surface density (in $\text{g}/\text{cm}^2$), and the y-axis indicates the number of simulations that produced a most massive planet with the mass shown on the x-axis. In general, higher surface densities lead to the formation of more massive planets. Planet masses are given in Earth masses ($\text{M}_{\oplus}$). }
    \label{fig:density_hist}
\end{figure}

Notably, the absence of a turnover point or flattening in the high-density regime suggests that the relative velocities excited between planetary embryos are low enough to allow growth even under crowded conditions. As shown in Figure~\ref{fig:density_hist}, formation {of the most massive planets} tends to favor the high density regime. 

This result underscores the importance of the initial mass budget in the inner disk. In systems with higher solid area densities, embryos experience more frequent and massive collisions, facilitating the growth of planets. This trend holds across a wide range of physically realistic area densities, suggesting that the outcome is robust over varying initial conditions. In the highest density cases, many of the resulting planets reaches sizes close to or above $1 M_{\oplus}$, placing them within the detectable range of current transit and radial velocity surveys.

\subsection{Effect of the Hot Jupiter Orbital Position}
Figure~\ref{fig:spacing_plot} illustrates the mass of the largest planet formed as a function of the hot Jupiter's orbital distance, varied from 0.03 to 0.10 au, as computed from our second suite of simulations. {We find a clear positive correlation, systems with close-in hot Jupiters begin with small embryos ($\sim2.4\times10^{-2}\text\ {M}_{\oplus}$ at $a_J \sim 0.03$ au) that grow to nearly an Earth mass through frequent mergers, suggesting that the hot Jupiter's dynamical influence enhances collisional activity. Contrastingly, when the hot Jupiter is farther from the star ($a_J\sim0.1$au), our simulated embryos are already much more massive ($\sim0.63\text{M}_{\oplus}$) and experience fewer mergers, so their final sizes, while larger, represent a relatively more modest growth beyond their initial masses. }

\begin{figure}
    \centering
    \includegraphics[width=1\linewidth]{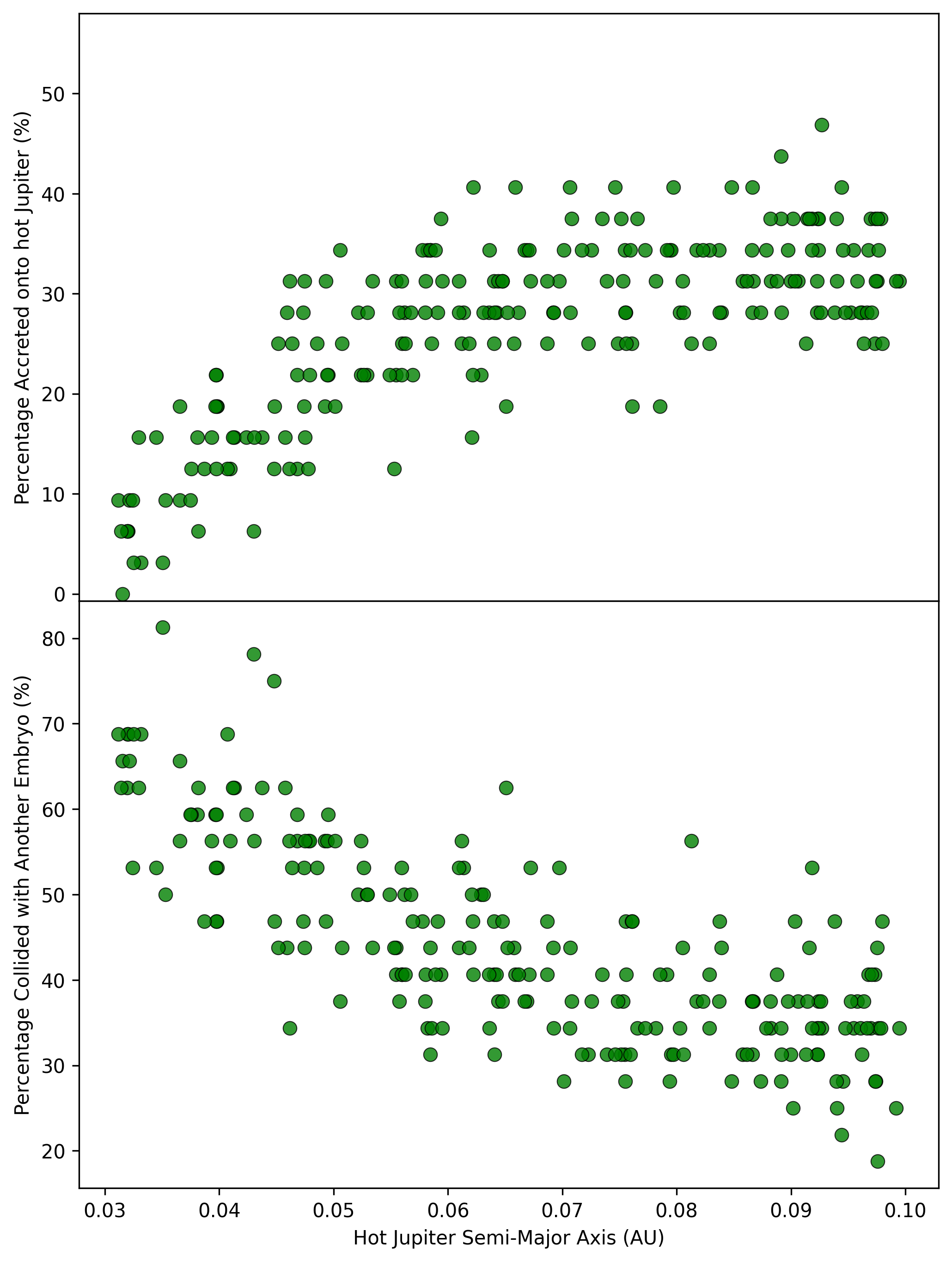}
    \caption{{The major outcomes of embryo interactions as a function of the hot Jupiter's semi-major axis. Alternate outcomes are significantly less common ({across} all simulations, {$\sim0.7\%$ accrete onto the star, $\sim 0.1\%$ are ejected from the system entirely and $\sim15\% $ are merger products or non-interacting embryos that survive until the end of the integration).} \textit{Bottom panel:} The percentage of embryos that merged {into another embryo}. \textit{Top panel:} The percentage of embryos that were accreted onto the hot Jupiter. We find that most solid material remains dynamically bound and redistributed within the inner system rather than being removed entirely.}}
    \label{fig:survival_plot}
\end{figure}

{In addition to influencing the maximum possible interior planet mass, the hot Jupiter's orbital position also strongly shapes the dynamical outcomes of planetary embryos. We classify embryos that persist until the end of the integration, {both non-interacting and merger products}, as {``survived''} embryos. In contrast, embryos considered lost to the system are those that are either dynamically ejected or accreted by the star, whereas accretion onto the hot Jupiter {is} treated as mass redistribution within the system. Figure~\ref{fig:survival_plot} summarizes the distribution of embryo survival outcomes across our simulations. We find that the vast majority of embryos remain in the system, with only a very small fraction being lost entirely. {Across all our simulations, $\sim 0.7\%$} of {interactions} resulted in stellar accretion and {$\sim 0.1\%$ in complete ejection. In our simulations, the dominant processes shaping the final system architecture are mergers rather than embryo loss.}}


{Taken together with Figure ~\ref{fig:spacing_plot}, these results highlight a key conclusion: embryos at close-in separations are typically not removed from the system, but rather have their mass redistributed through collisions and mergers. At shorter orbital radii ($a_J \lesssim0.06$ au), the dominant outcome is embryo-embryo collisions, which both reduce the number of distinct objects and produce planets up to $\sim1\text{M}_{\oplus}$. At larger separations ($a_J\gtrsim0.06$ au), accretion onto the hot Jupiter becomes increasingly dominant, while surviving embryos persisting as distinct bodies exhibit a small fractional growth (although the final particle masses, shown in Figure \ref{fig:spacing_plot}, are more massive since the initial particle sizes in our simulations were larger). Overall, the system's architecture is shaped primarily by the redistribution of embryo mass through mergers, accretion onto the hot Jupiter, and surviving embryos, with embryo loss playing only a minor role.}

\vspace{4mm}
\section{Discussion\label{sec:discussion}}
In this work, we consider the in situ formation of hot Jupiters and their interior companions. In both cases, `in situ' means that the planets assemble to their final masses in roughly their final orbital positions. However, we assume that the building blocks of both planet types must form further out in the protoplanetary disk and then be delivered inward via disk migration. For the hot Jupiters, the required building block is a super-Earth-sized core, which can then undergo runaway gas accretion at its final orbital position. For the inner companions, the required building blocks are planetary embryos, which can grow into planets via collisional accretion.

Our results demonstrate that the formation of terrestrial planets interior to hot Jupiters depends on both the initial mass distribution in the inner disk and less sensitively on the orbital radius of the gas giant. Intuitively, higher area densities of planetary embryos dramatically enhance the likelihood of forming large terrestrial planets, as collision rates are elevated in these massive inner disks.

We find that the survival percentage of distinct embryos, {including merger products, behaves unremarkably} with the hot Jupiter’s orbital distance. At small separations, strong gravitational perturbations from the giant planet rapidly destabilize the inner disk, leading to frequent embryo mergers{, yet the percentage of survivors changes little}. As the hot Jupiter moves outwards, its influence is diminished {and the survivor fraction remains roughly unchanged}. This suggests that, {even} when the hot Jupiter is farther out, {embryos are not more likely to survive, but instead experience reduced collisional growth with fewer mergers. This is} consistent with the mass trend seen in Figure~\ref{fig:spacing_plot}

\subsection{The Warm Jupiter Connection}
In this work, we are primarily concerned with the formation of hot Jupiters and their interior companions. However, while hot Jupiters do not often have such interior companions, warm Jupiters are observed to have interior companions much more frequently \citep{Huang2016}. 
The proposed mechanism we consider in this work to form these interior companions involves planetary embryos being formed at radii that initially place them in dynamically decoupled locations in the disk, and then being funneled into highly interacting orbits via convergent migration driven by an exterior Jupiter. 
This mechanism would work for interior companions to both warm and hot Jupiters. Although our results did not specifically address how the more distant configurations of warm Jupiters might decrease the interaction frequency between planetary embryos, it is likely that a warm Jupiter would lead to a lower surface density of planetary embryos in the disk, resulting in less efficient planet formation.

The in situ formation mechanism for hot Jupiters and their interior companions is likely partially consistent with observed constraints on stellar obliquity \citep[where warm Jupiters tend to have aligned stellar obliquities;][]{Morgan2024} and parameter distributions. However, some subset of warm Jupiters exhibit larger eccentricities, which are consistent with tidal migration \citep{Petrovich2016} and may be further affected by additional perturbing bodies in their systems \citep{Mustill2017}. Therefore, in situ formation likely contributes to only a fraction of the full sample of warm Jupiters with companions.

\subsection{Caveats and Future Work}
One open question not addressed in the present work is the detailed dynamical pathway responsible for the short-scale migration of the proto-hot Jupiter core from its initial formation location to its final close-in orbit. 
Similarly, the mechanism that halts this migration and sets the final orbital period of the hot Jupiter is equally uncertain. 
The specific dynamics of this short-scale migration also govern whether planetary embryos interior to the migrating core survive or are destabilized during the migration process. Their survival is necessary to form the type of inner disk of planetary embryos explored in this study.

In this work, we assume that such a migration occurred and that it transported solid material inward, assembling a population of planetary embryos interior to the hot Jupiter's final orbit. 
However, previous work by \citet{Hallatt2020} has suggested that reconciling disk migration with observed gas giant parameters is not straightforward, potentially posing a challenge for mechanisms requiring specific outcomes of disk-migration \citep{Gan2024, Su2024}. {Similarly, depending on the exact migration parameters, the planetary embryos may attain {resonant} configurations during migration \citep{Mandell2007}, affecting their ability {to accrete} into larger planets.}

Another scenario not considered in this work is the possibility that a planetary companion forms prior to the short-scale disk migration of the hot Jupiter core and migrates inward in its fully assembled form. In this case, the inner companion would not form from in situ accretion of embryos but instead survive the migration process as an intact planet, ultimately becoming an interior companion to the hot Jupiter. However, the argument presented in the top panel of Figure \ref{fig:area_density_plot}, that the orbital radius at which the hot Jupiter core formed can be determined by the total amount of interior material,  would still hold in this case, but potentially with different scalings than we find in this work.

{We also note that the mechanism explored here cannot account for all observed hot Jupiter companions. In particular, several known systems host Neptune-sized interior planets with densities around $\rho \approx 2-3$ g/cm$^{3}$ \citep{Korth2023, Korth_2024}, which likely to retain H/He envelopes. Such planets must have formed early, while the gas disk was still present \citep[the scenario considered in][]{Poon2021}, and therefore cannot be produced by the post-gas-disk scenario we consider. Measuring the bulk densities of such companions will be necessary to distinguish which systems are compatible with the in situ assembly mechanism studied here and which require alternative formation pathways.}

{As pointed out in \citet{Poon2021}, simulations of in situ formation of hot Jupiters over-predict the number of hot Jupiters assumed to have inner companions. Our results (Figure \ref{fig:area_density_plot}) also show that inner companions will be observable for a wide range of formation assumptions, and \citet{MacLean2025} showed that such companions, once formed, should not attain enough mutual inclination with the hot Jupiter to not be seen in transit. This is at odds with the observed {rarity} of inner companions to hot Jupiters, suggesting that the in situ formation mechanism is not dominant in {forming} hot Jupiters as a population.}

Additionally, in this work, we do not consider the subsequent dynamical {evolution of the planetary system after our simulations conclude}. Secular eccentricity excitation, resonant dynamics, and tidal evolution will cause further changes in orbital architectures on timescales much longer than that considered in this work \citep[see for example,][]{Wu2023, He2024, Wang2025}.

\section{Conclusion} \label{sec:conclusion}

In this study, we explore the potential for in situ formation of terrestrial planets interior to hot Jupiters, using N-body simulations to investigate the effects of solid area density and the orbital position of the hot Jupiter. Our results indicate that the formation of larger planets is facilitated by higher surface densities of planetary embryos, which enhances the rate of collisional accretion. We also find that the position of the hot Jupiter plays a role in determining the size of the formed planets, but this factor is not as important as the disk surface density. These findings suggest that in situ formation of inner planets is a plausible mechanism, particularly under conditions of high embryo density and a moderately close-in hot Jupiter. However, the interaction between short-range migration, planetary embryo dynamics, and the overall disk structure will continue to require further investigation to fully understand the variety of planetary systems observed today.

\section*{Acknowledgments}
We would like to thank the CHTC network and their team, especially Andrew Owens, at UW-Madison for their resources and assistance. 
We thank Dr. Mariah MacDonald, Adam Distler, and Pulkit Mohata for useful conversations. We thank Fred Adams for formative discussions that directly inspired the inception and direction of this project. {We thank the referee for substantially useful comments.}

\software{
\texttt{matplotlib} \citep{Hunter:2007},
\texttt{pandas} \citep{mckinney-proc-scipy-2010, the_pandas_development_team_2024_13819579},
\texttt{REBOUND} \citep{Rein2012}
\texttt{TRACE} \citep{Lu2024}, Astropy \citep{astropy:2013, astropy:2018, astropy:2022}, \texttt{scipy} \citep{2020SciPy-NMeth}}

\facilities{CHTC \citep{CHTC}, Exoplanet Archive \citep{Christiansen2025}}

\bibliography{references}
\end{document}